\documentstyle[12pt]{article}

\newcommand{\be}{\begin{equation}}
\newcommand{\ee}{\end{equation}}
\newcommand{\zl}{<-{1\over 2}{\bf Z}\mid}
\newcommand{\zr}{\mid {1\over 2}{\bf Z}>}
\newcommand{\zz}{<-{1\over 2}{\bf Z}\mid {1\over 2}{\bf Z}>}

\newcommand{\al}{\alpha}
\newcommand{\bary}{\begin{eqnarray}}
\newcommand{\eary}{\end{eqnarray}}
\begin{document}
\begin{titlepage}
\title{Charge Radii of Hyperons}
{\author{Sarira Sahu$^{1}$\\
Departament de F\'{\i}sica Te\`{o}rica and I.F.I.C.\\
Centre Mixt Universitat de Val\`{e}ncia -- C.S.I.C.\\
E-46100 Burjassot (Val\`{e}ncia), Spain.}}
\date{ }
\footnotetext[1]{email:sahu@evalvx.ific.uv.es}
\maketitle
\thispagestyle{empty}
\begin{abstract}
The momentum projected SU(3) Chiral Color Dielectric Model (CCDM)
is employed to
calculate the charge root mean square radii and charge distributions of
hyperons. We compare our result with Skyrme, MIT bag and Cloudy Bag model
results. The charge distribution of $\Lambda$ in CCDM is similar to that
of Skyrme Model prediction.
\end{abstract}
\end{titlepage}
\vfill
\eject
Apart from the proton and neutron charge radii, no other baryon charge radius
has so far been measured. So there are predictions of the hyperon charge
radii in different models. In all the isospin symmetric quark models
the quark contribution to neutron charge radius is zero, although the
experimental value of neutron charge radius is -0.35 $fm$. In SU(2)
Cloudy Bag Model (CBM) picture\cite{tho},
the neutron has a $p\pi^-$ component, which
has a positive charge distribution inside and negative charge distribution
outside. So the net charge distribution seen by the electromagnetic probe
is negative, which consequently gives negative charge radius to neutron.
Also the pion cloud in CBM contributes to all other static properties
of the baryon. The chiral SU(2) version of CBM\cite{tho},
Friedberg-Lee soliton model\cite{lub} and
color dielectric models\cite{lee,bir,sah2}
have been studied extensively. All these
models predict the hardonic static properties successfully. These models
have been extended to include SU(3) chiral symmetry to study the
contribution of pseudoscalar meson octet, although this symmetry is
broken due to the different masses of pion, kaon and eta
mesons\cite{sah3}.

    In this letter we have calculated the charge root mean square radii
and the electric charge distribution of hyperons in SU(3) CCDM. Momentum
projection technique of Peierls and Yoccoz is used to over come the problem of
center of mass motion. We compare our result with the other models. It
shows that the charge distribution of $\Lambda$ in CCDM is similar to
Skyrme model prediction.

Upto the first power in $1/f_\phi$ the CCDM lagrangian is given by\cite{sah3},
\bary
{\cal L} &=&
\sum_i{\bar\psi}_i (x)\Big [i\gamma^{\mu}\partial_{\mu}
- (m_{su}+{m_u\over \chi(x)}
(1+{i\over f_{\phi}}\gamma_5 \lambda . \phi(x)))
-{1\over 2} g_s \gamma^{\mu}\lambda_a A_{\mu}^a (x)\Big]\psi_i (x)
\nonumber\\
& &-{1\over 4}\epsilon(\chi) F^a_{\mu\nu}(x)F^{a,\mu\nu}(x)
+{1\over 2}\sigma_v (\partial_{\mu} \chi (x))^2
-U(\chi(x))
\nonumber\\
& &+{1\over 2} (\partial_{\mu} {\phi(x)})^2
-{1\over 2}m^2_{\phi}{{\phi}(x)}^2,
\eary
where $\psi(x)$, $\chi(x)$, $\phi(x)$ and $A_\mu^a(x)$ are
the effective quark,
color dielectric, pseudoscalar meson and gluon fields respectively,
$f_\phi$ is the meson decay constant, $m_{\phi}$ is the octet meson mass and
$\al_s={g^2_s \over {4\pi}}$ is the strong coupling constant.
The effective quark mass,
$m_{su}+{m_u \over \chi}$ is assumed to break SU(3) flavor symmetry
by assuming that $m_{su}$ is zero for u and d quark and nonzero for s quark.
The effective gauge field $A^a_{\mu}$ interacts with the dielectric
field $\chi$, through a dielectric functional $\epsilon (\chi)$ = $\chi^4$.
The self interaction of the dielectric field is assumed to
be of the form,
\be
U(\chi)=B\Big [\al\chi^2-2(\al-2)\chi^3+(\al-3)\chi^4\Big ]
\ee
with U having an absolute minimum at $\chi=0$ and a secondary minimum
at $\chi~=~1$. The quantity $m_{GB}~=~\sqrt{ {2B\al} \over {\sigma^2_v} }$
can be identified with the mass of the dielectric field and is often
interpretted as the glueball mass.
In the physical vacuum, the
dielectric field vanishes and the effective quark mass term
$m_u\over \chi$ goes to
infinity, thus confining the quark by an infinite potential well.
Similarly the dielectric functional $\epsilon (\chi)$ goes to zero in
the physical vacuum and gluon field vanishes. Thus the quarks and the gluos
exist only in the region where $\chi$ $>$ 0.

Scaling and non-scaling versions of the CDM has been used earlier to
calculate the baryonic properties\cite{sah1}. In scaling model the strange
quark
mass is given by $m_s=(m_u+m_{su})/\chi(0)$ and in the non-scaling
model $m_s=m_{su}+m_u/\chi(0)$. Thus the strange quark mass in both
the cases are different. Apart from that the confinement of strange
baryons in scaling model are different from the non-strange baryons.
On the other hand the confinement is same for strange and non-strange
baryons in non-scaling model and it is similar to the MIT bag model.
Here we work with the non-scaling model. The non-scaling model has also
been used
earlier to calculate the baryonic properties in SU(3) chiral CDM\cite{sah3}.
As the scalar field accounts for the long range
order effect of the QCD vacuum, we treat the gluon contribution
perturbatively. We consider only one gluon exchange interaction,
because more than one gluon exchange may give rise to color singlet quanta
i.e. $\chi$, which is already present in the model.

The mean-field solution of the scalar field obtained in CDM is
localised and contain spurious cente of mass (CM) energy. It also
smears out charge distribution, for example increasing the charge
radii of the baryons. Various techniques have been developed to avoid
this CM motion. The momentum projection technique of Peierls and
Yoccoz has been used extensively\cite{lub,lee,sah4}
to obtained an approximately correct
picture of hadron which is an eigenstate of the momentum. With the
help of the projected state the baryonic static properties can be
calculated.

A baryon state with finite momentum $\bf P$ is given as\cite{lub}
\be
\mid {\bf P}> =\int d^3x e^{i{\bf P.X}}\mid X>,
\ee
where $\mid X>$ is a localised baryon state centred at ${\bf X}$ and
is given by
\be
\mid X> = exp \Big [\int d^3k {\sqrt{\omega_k\over 2}}f_k(X)a^{\dagger}_k
\Big ]
b_1^{\dagger}(X)b_2^{\dagger}(X)b_3^{\dagger}(X)\mid 0>.
\ee
The exponential term in the above equation is the coherent dielectric
field (scalar field) and $a^{\dagger}_k$ is the creation operator for
the dielectric field. Similarly $b_i^{\dagger}$ are the creation
operators for the quark fields.

As the physical baryon in this picture is a system of three quarks surrounded
by a meson cloud, the state of a baryon A is expressed as\cite{tho}
\be
|A>~=~\sqrt{P_A}\Big \{~1~+~(m_A-H_0)^{-1}H_{int}~\Big \}|A_0>.
\label{prob}
\ee
Here $P_A$ is the probability of finding a bare baryon state $|A_0>$
with bare mass $m_{A_0}$ in the baryon A. $H_0$ is the noninteracting
Hamiltonian which includes  quark and dielectric field Hamiltonian and
free meson Hamiltonian . $H_{int}$ is the interaction
Hamiltonian, which includes meson-quark interaction. The second term in
eq(\ref{prob}) generates the  meson cloud around the baryon A.
The photon couples with work as well as the meson. So
in general both quark and meson contribute to the baryon properties.
 The core
(quark) contribution comes from the coupling of photon to the quarks which
also includes the term when one meson is "in the air"\cite{tho}.
The meson contribution comes
from coupling of a photon to the meson. Due to the meson cloud around the
baryon
the electromagnetic current of baryon has contribution
from the quark sector as well as the meson sector.
So the total baryon electromagnetic current is
\bary
J^{em}_{\mu}({\bf x}) &=&
\sum_i Q_i{\bar\psi_i}({\bf x})\gamma_{\mu}\psi_i({\bf x})
+ie\big[{\phi^{\dagger}(x)\partial_{\mu}\phi(x)}
-\phi (x){\partial_{\mu}\phi^{\dagger}}(x)\big]
\nonumber\\
&=& J^{em}_{\mu,q}(x)+J^{em}_{\mu,\phi}(x)
\eary
where the charge operator is given as
\be
Q = e({1\over 2}\lambda_3 + {1\over {2\sqrt 3}}\lambda_8).
\ee
The charge mean square radius $<r^2>$ of the baryon is
obtained from the matrix element of the charge density with respact to
the baryon state. The matrix element of the charge density can be written as
\be
{{<{1\over 2} P\mid J^{em}_0(0)\mid -{1\over 2} P>}
\over
{<{1\over 2} P\mid {1\over 2} P>}}
=
{
{\int d^3Zd^3r e^{iP.r}\zl J^{em}_o(r)\zr}
\over{\int d^3Z\zz}}
\label{rms}
\ee
By expanding the RHS of eq(\ref{rms}) to second order in $\bf P$ we
get\cite{sah3}
\be
{{\int d^3Zd^3r e^{iP.r}\zl J^{em}_0(r)\zr}
\over{\int d^3Z\zz}}
=1-{<r^2>\over 6}P^2 + {\cal O} (P^4),
\ee
where $<r^2>$ is the charge radius.
The quark part of the charge square radius is calculated using the
momentum projected baryon states as described above and the meson cloud
contribution is calculated from the
electric formfactor $G_E(q^2)$ (meson part only) of the
baryon, which is given as $<r^2>=-6{\partial\over{\partial q^2}}
G_E(q^2)_{q^2=0}$\cite{tho,sah3}.

The parameters in the CCDM are $m_{GB}$, $m_u$, $m_{su}$, $\al_s$,
B, $f_\phi$ and $\al$. Our calculations\cite{sah3} show that $\al$,
the parameter
that which determines the height of the maximum of $U({\chi})$ between
two minima at $\chi=0$ and $\chi=1$ does not play an important role
in the calculation. Therefore we choose $\al=36$ throughout. Also,
the meson-quark
coupling constant $f_\phi$ has been chosen to be 93 MeV, the pion decay
constant. The rest of the parameters are varied to fit the properties
of octate and decuplet of baryons. We find that the masses of these
baryons can be fitted, to a very good accuracy, for a family of parameter
sets\cite{sah3}. In particular, we find that the glueball mass can vary from
about 1 GeV to 3 GeV. However, the other static properties
of hadrons are best reproduced for $m_{GB}$ of about 1 GeV. These
generally become smaller as $m_{GB}$ is increased.
The result for a particular set of parameter is given in table 1. For
the above set we obtain resonablely good fit for baryon masses and magnetic
moments.

   As we are interested in the total charge radii as well as their charge
distributions, we do not calculate the quark and the meson contributions
seperately. We have all ready shown in a previous work
\cite{sah5} that the meson exchange contribution to nucleon charge
radii is small. Also the kaon and eta contributions are negligablely
small. Apart from that as the number of strange quark increases in the
baryon, the pion coupling decrases and is zero for $\Omega^-$. We observed
that compared to quark contribution, the meson contribution is very small
except for nucleon.

Table 1 shows that the proton charge radius in CCDM is close to the result
obtained by cloudy bag model (CBM)\cite{tho} and the Skyrme model\cite{kun}
and also very close to
the observed value (0.82 fm).
For all, isospin symmetric quark models,
the quark contribution to neutron charge radius is zero. But in chiral
quark models the meson cloud and the meson exchange interactions
give negative contribution to the neutron charge radius and we find it to be
$-0.284$ fm which is very close to the observed value.

Fig. 1 shows the charge density of $\Lambda$ in CCDM.
The strangeness density
is negative and is localised inside the $\Lambda$.
Thus the electric charge density of $\Lambda$ is negative inside and positive
out side due to the light quarks contribution. Similar charge distribution is
also obtained in Skyrme model\cite{kun}.
The charge radius of $\Lambda$ in our
model is very close to that of the Skyrme model\cite{kun}
one as shown in table 1.
As the cloud contribution is proportional to the third component of the
isospin, it is zero for $\Lambda$ and $\Sigma$. Only the exchange term
contribute to the charge radius of $\Lambda$ and $\Sigma$ and it is very samll.
The mesonic contribution for $\Lambda$ and $\Sigma^0$ are almost same. Apart
from that the quark contribution to charge radii of both these baryons are
same. So in CCDM the charge radii of $\Lambda$ and $\Sigma^0$ and their
corresponding charge densities are almost same.
We find the charge rms radius of $\Lambda$ to be
larger than precicted by MIT bag model and the CBM one.

The charge distribution for $\Sigma^+$ and
$\Sigma^-$ are positive and negative respectively as shown in fig. 2.
In CCDM the magnitude of charge radii of $\Sigma^+$ and $\Sigma^-$ are large
compared to other models.
The $\Xi^0$ charge density in fig. 3 shows that, the core has negative charge
distribution due to the strange quark contribution followed by a positive
charge distribution towards the surface, like the $\Lambda$ charge distribution
as shown in fig. 1.
The $\Omega^-$ charge density is negative and the
magnitude of its rms radius is also higher than the corresponding results
obtained for other models.

We have calculated the charge rms radii of hyperons in SU(3) CCDM.
We obtain the correct rms radius of proton
and also the neutron charge radius is very close to the observed valus.
It is observed that the rms radii of $\Lambda$ and $\Sigma$ in CCDM
are very close to the Skyrme model prediction and larger than the
corresponding results
obtained in MIT bag model and CBM. The charge densities of all the
neutral hyperons show a negative charge density inside the core and a
positive charge density on the surface. Compared to other models, in CCDM
the magnitude of the rms radii of charged hyperons are large except $\Xi^-$.

It is a pleasure to thank Prof. E. Oset and Prof. V. Vento for many helpful
discussions. The author is thankful to the Ministerio de Educaci\'on y
Ciencia, Goverment of Spain for  the financial support.

\begin{table}
\caption{The charge root mean square radii of the baryons in CCDM
with $m_u=m_d$ = 105 MeV, $B^{1/4}$ = 94.5 MeV, $m_{GB}$ = 1050 MeV and
$m_s$ = 318 MeV is compared with the results of
Skyrme with $m_{\pi}$ = 138 MeV and
MIT bag model with $m_u$ = $m_d$ = 0 and $m_s$ = 279 MeV and the CBM.}
\vspace {0.5in}
\begin{tabular}{|c|c|c|c|c|}
\hline
\multicolumn{1}{|c|}{$\it particle$}&
\multicolumn{1}{|c|}{$<r>^{CCDM}$}&
\multicolumn{1}{|c|}{$<r>^{Skyrme}$}&
\multicolumn{1}{|c|}{$<r>^{MIT}$}&
\multicolumn{1}{|c|}{$<r>^{CBM}$}\\
\hline
\hline
p          & 0.814   & 0.88    & 0.73    & 0.84\\
n          & $-0.284$     & $-0.55$ & 0.0     & $-0.35$\\
$\Lambda$  & 0.287   & 0.33    & 0.16    & 0.008\\
$\Sigma^+$ & 0.841    & 0.98    & 0.75    & 0.75\\
$\Sigma^0$ & 0.286   & 0.33    & 0.16    & 0.14\\
$\Sigma^-$ & $-0.737$ & $-0.87$ & $-0.71$ & $-0.76$\\
$\Xi^0$    & 0.554    & 0.47    & 0.23    & 0.15\\
$\Xi^-$    & $-0.593$ & $-0.51$ & $-0.69$ & $-0.72$\\
$\Omega^-$ & $-0.86$ & $-0.38$ & $-0.67$ & $-0.75$\\
\hline
\end{tabular}
\end{table}
\vfill
\eject

\vfill
\eject
\noindent{Figure Captions}

\noindent{Figure 1}

\noindent {The charge density of $\Lambda$
is plotted as function of radius.}

\noindent{Figure 2}

\noindent {The charge densities of  $\Sigma^+$ and
$\Sigma^-$ are plotted as functions of radius.}

\noindent{Figure 3}

\noindent {The charge densities of $\Xi^0$ and $\Xi^-$
are plotted as functions of radius.}
\vfill
\eject
\end{document}